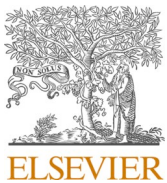
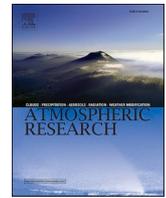

# On the impact of initial conditions in the forecast of Hurricane Leslie extratropical transition

M. López-Reyes [a,b], J.J. Gónzalez-Alemán [c], M. Sastre [a], D. Insua-Costa [d,e], P. Bolgiani [a], M. L. Martín [f,g,*]

[a] *Department of Earth Physics and Astrophysics, Faculty of Physics, Complutense University, Madrid, Spain*
[b] *Astronomy and Meteorology Institute, Physics Department, University of Guadalajara, Guadalajara, Mexico*
[c] *Agencia Estatal de Meteorología (AEMET), Department of Development and Applications, Madrid, Spain*
[d] *Hydro-Climate Extremes Lab (H-CEL), Ghent University, Ghent, Belgium*
[e] *CRETUS, Non-linear Physics Group, Universidade de Santiago de Compostela, Galicia, Spain*
[f] *Department of Applied Mathematics, Faculty of Computer Engineering, University of Valladolid, Spain*
[g] *Institute of Interdisciplinary Mathematics (IMI), Complutense University of Madrid, Madrid, Spain*

ABSTRACT

Hurricane Leslie (2018) was a non-tropical system that lasted for a long time undergoing several transitions between tropical and extratropical states. Its trajectory was highly uncertain and difficult to predict. Here the extratropical transition of Leslie is simulated using the Model for Prediction Across Scales (MPAS) with two different sets of initial conditions (IC): the operational analysis of the Integrate Forecast System (IFS) and the Global Forecast System (GFS).

Discrepancies in Leslie position are found in the IC patterns, and in the intensity and amplitude of the dorsal-trough system in which Leslie is found. Differences are identified both in the geopotential height at 300 hPa and the geopotential thickness. Potential temperature in the dynamic tropopause shows a broader, more intense trough displaced western when using the IC-IFS. The IC-IFS simulation shows lesser trajectory errors but wind speed overestimation than the IC-GFS one. The complex situation of the extratropical transition, where Leslie interacts with a trough, increases the uncertainty associated with the intensification process.

The disparities observed in the simulations are attributed to inaccuracies in generating the ICs. Both ICs generate different atmospheric configurations when propagated in time. Results suggest that during an extratropical transition in a highly baroclinic atmosphere, the IFS model's data assimilation method produced a more precise analysis than GFS due to the greater number of observations assimilated by the IFS, the greater spatial resolution of the model and the continuous adjustment of the simulations with the field of observations.

## 1. Introduction

In 2018, Hurricane Leslie developed from a system of non-tropical origin, exhibiting a remarkably long-life cycle, and undergoing multiple tropical and extratropical transitions. It has been one of the few systems with tropical characteristics and of greater intensity that have impacted the Iberian Peninsula (Stewart, 2018). As it traversed through Spain and France during its extratropical depression phase, the system Leslie interacted with a frontal system, resulting in the intensification of precipitation in western Spain and southern France. This event led to an accumulated maximum precipitation of nearly 300 mm in <24 h and wind speed values exceeding her than 100 km/h (Mandement and Caumont, 2021), which caused flash floods, leaving 13 people dead and costly property damage.

In the final phase of Leslie's life cycle, it underwent an extratropical transition (ET), favored by the synoptic conditions (Pasch and Roberts, 2019). The system was located east of a trough axis in the region of upper-level divergence, so a baroclinic atmosphere supported the change in the cyclone structure without losing much intensity (Sadler, 1976; Pérez-Alarcón et al., 2023). It is worth noting the upstream presence of another ET, associated with the major Hurricane Michael, interacting downstream with the Leslie environment (Beven II et al., 2019). In the operational forecasts, it was difficult to predict Leslie's trajectory and intensity since the models showed high uncertainty, even

---

within a few hours of the initializations.

During an ET, a storm may intensify as it begins to tap baroclinic energy in addition to the energy derived from the thermodynamic imbalance between the atmosphere and the underlying ocean (Evans et al., 2017; Bieli et al., 2019; Quitián-Hernández et al., 2016). During and after an ET, the system frequently accelerates its translational motion, resulting in heavy precipitation and strong redistributed winds that endanger coastal regions and maritime activities (Jones et al., 2003). According to Komaromi and Doyle (2018) and Pantillon et al. (2013), when an ET effectively undergoes, the tropical cyclone (TC) must be in phase with a trough in the middle or upper troposphere, so that the ascent region coincides with the location of the vortex. Upward movements are generated by the trough in conjunction with the TC's own convection. Also, ET intensification occurs when the cyclone matches with a jet streak or with a local intensification of the extratropical jet stream (Pantillon et al., 2013), such that the divergence region at the entrance to the right of the jet streak coincides with the cyclone on the surface (Maddison et al., 2020).

As mentioned before, the positioning of the TC relative to the trough plays a pivotal role to an ET and its intensification (Anwender et al., 2008). Previous studies such as Keller et al. (2019) and McTaggart-Cowan et al. (2003, 2004) showed that the presence of an additional ET or any other process characterized by vigorous diabatic heating upstream can induce the release of latent heat, resulting in the reorganization of isentropes. This will lead to a decrease in the upper-level potential vorticity (PV) and an amplification of the ridge. Such dynamics establish a preconditioning effect on the intensity and amplitude of the trough downstream from the ridge, adding uncertainty to the behavior of the air flow.

The use of numerical weather prediction models has shown it is hard to forecast ETs, especially with lead times >36 h (Evans et al., 2017). Veren et al. (2009) found that forecast errors in the intensity and trajectory of TCs having an ET are significantly higher compared to those systems that are already extratropical. For example, Munsell and Zhang (2014), Munsell et al., 2015) show high variability in track and intensity forecasts for hurricanes Nadine [2012] and Sandy [2012], even using sophisticated data assimilation methods. This variability is associated with diabatic processes and their influence on the amplification and interactions of the crest-valley flow (Bassill, 2014; Torn et al., 2015). In the same way, Scheck et al. (2011) propose the concept of bifurcation points that consists of a theory on the influence of tropical and extratropical characteristics on the sensitivity of the models for trajectory and intensity forecast during an ET. An example of this bifurcation sensitivity can be seen in small errors in the representation of the interaction of the trough with the TC vortex resulting on a remarkably different evolution in the interaction, i.e., small variations in the initial situation result on an amplification of errors (Evans et al., 2017).

Since the trajectory and intensity forecast of the Hurricane Leslie with numerical models entailed high uncertainty, the current study deals with the Leslie ET occurred in the final cycle life of the system using simulations of the Model for Prediction Across Scales (MPAS) from two different initial conditions (ICs) of the global models. Here, the Global Forecast System (GFS) and the Integrated Forecasting System (IFS) of the European Center for Medium-Range Weather Forecasts (ECMWF) are selected. Given Leslie's high uncertainty and taking it as an example of a highly unpredictable atmosphere, the primary goal of this study is to investigate possible differences between the MPAS simulations of Leslie's ending stage resulted from these ICs, in addition to learn about the impact the ICs have on the prediction of a system with such characteristics.

This paper is organized as follows. Section 2 includes the description of the model, the data and the methodology used. In section 3, the results and discussions are presented and finally, the conclusions and perspectives are provided in section 4.

**Table 1**
Parameterization schemes included in each of the default MPAS physics suites (schemes and acronyms can be revised in Duda et al., 2019).

| Parameterization | Mesoscale reference | Convection-permitting |
|---|---|---|
| Microphysics | WSM-6 | Thompson |
| Convection | New Tiedtke | Grell Freitas |
| PBL | YSU | Mynn |
| GWDO | YSU GWDO | YSU GWDO |
| Longwave radiation | RRTMG | RRTMG |
| Shortwave radiation | RRTMG | RRTMG |
| Cloud | Fraction | Fraction |
| Surface layer | Monin-Obukhov | Mynn |
| Land surface model | Noah | Noah |

## 2. Model set-up and methodology

The MPAS model (version 7.0; Skamarock et al., 2012), developed primarily by the National Center for Atmospheric Research (NCAR), is used in this study. MPAS uses a similar dynamic nucleus and same physical parameterization as the Weather Research and Forecast model (WRF; Rauscher and Ringler, 2014; Skamarock et al., 2008). The MPAS is a three-dimensional, limited-area, finite-differences, non-hydrostatic model, used mainly in atmospheric research.

The horizontal discretization is based on a non-structured Voronoi center type mesh grid-C of the state variables and in the edges the dynamic variables, varying from a global mesh 60 km to 3 km, around the location of Leslie in the Atlantic Ocean, off the coast of the Iberian Peninsula. MPAS uses different parametrization schemes for physical processes in the atmosphere. However, the model proposes two predefined sets of physical parametrizations to facilitate compatibility: the mesoscale scheme and the convective parameterization scheme (Duda et al., 2019). Table 1 shows the details of the corresponding parametrizations for each physics suite. Here, convection-permitting is used.

Two different ICs are used to feed the simulations of the Leslie system with the MPAS model: the GFS analysis from the National Centers for Environmental Prediction (NCEP) and the IFS analysis of the National Meteorological Archival and Retrieval System of the ECMWF. Outputs obtained for the hurricane Leslie using the MPAS model with those initial conditions are hereafter referred to as MPAS-G and MPAS-I, respectively. The analyzes are carried out with the operational versions of the global models in 2018, cycle 45r1 for the ECMWF-IFS with a horizontal resolution of 9 km and 137 vertical levels (European Center for Medium-Range Weather Forecasts, 2023b). More information about ECMWF-IFS model can be found in European Center for Medium-Range Weather Forecasts (2023b). The v14.0.0 version is selected for the GFS with a horizontal resolution of 0.25° (approximately 26 km) and 64 vertical levels (National Center Environment Information, 2023). More information about GFS model is in National Center Environmental Prediction (2021). The period of integration in the model simulation begins on October 13, 2018, at 00:00 UTC and extends through October 14, 2018, at 18:00 UTC with a temporal output of 3 h, covering enough margin to study the entire ET of the Leslie system. The model ability to simulate the event is evaluated using the trajectory and intensity data. A Python algorithm (tracking) is used to find the minimum central pressure from the simulated mean sea level pressure (MSLP), and another algorithm is used to determine the location of the point where the wind speed module at 10 m ($V$) is maximum within the cyclone's radius of maximum winds.

Various skill scores are used to evaluate the model simulations against the US National Hurricane Center (NHC) best track and intensity observations. In this study, the performance of the MPAS-I and MPAS-G is assessed for the trajectory and intensity of the Leslie system, calculating each score as a function of performance throughout the simulation. The metrics here considered are the root mean square error (RMSE) and the standard deviation (STDEV), defined as:

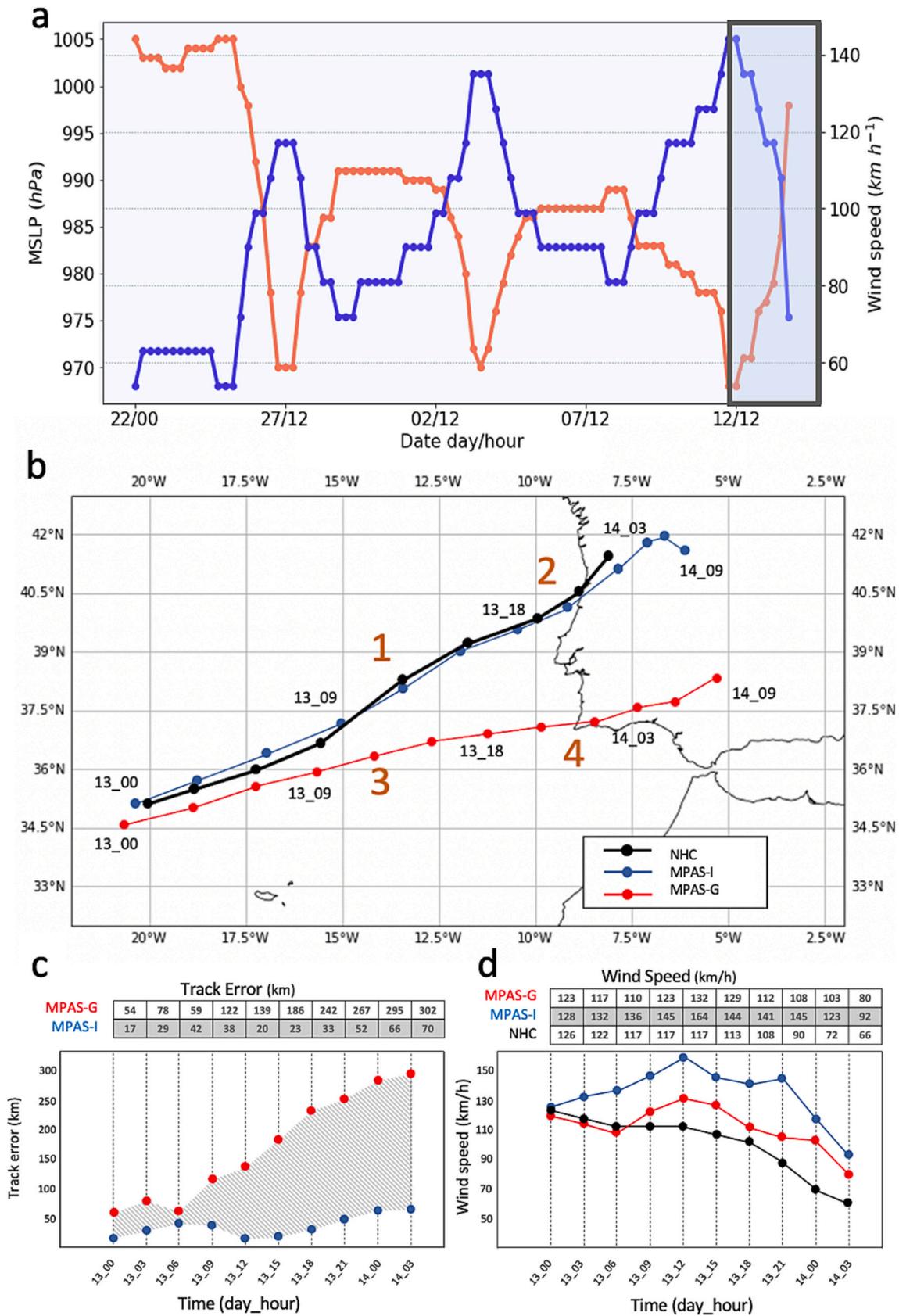

**Fig. 1.** (a) Evolution of Leslie's wind speed (blue) mean and MSLP (orange) throughout its life cycle. Data from Pasch and Roberts (2019); (b) Leslie trajectory of the two simulations and NHC official track, (c) Trajectory errors (km) of both simulations respect NHC best track and (d) Evolution of intensity (km/h) for the two simulations and official NHC intensity. (For interpretation of the references to colour in this figure legend, the reader is referred to the web version of this article.)

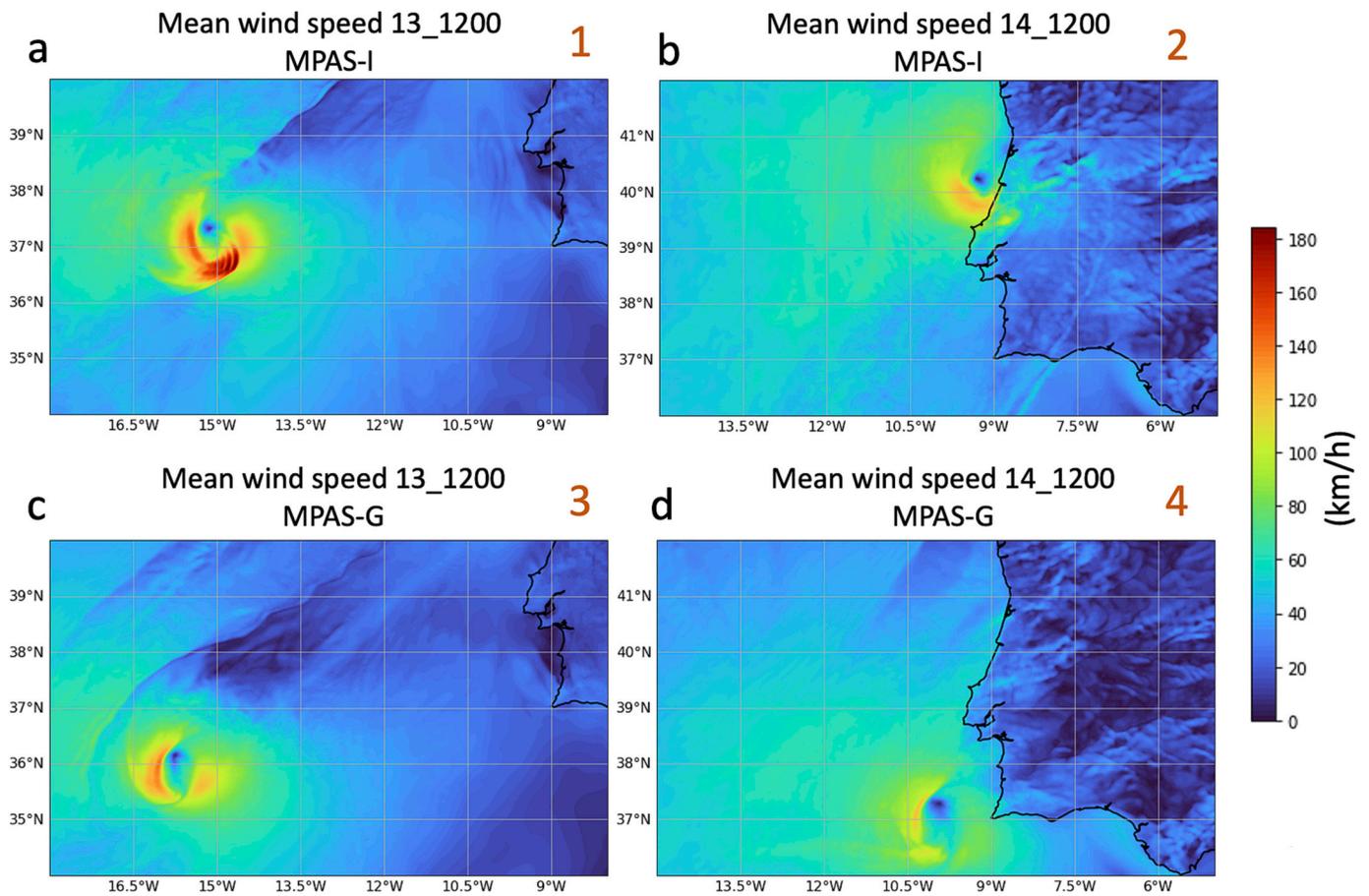

**Fig. 2.** Mean wind speed on 13 October 2018 for (a) MPAS-I; (c) MPAS-G; on 14 October 2018 for (b) MPAS-I; (d) MPAS-G, at 12:00 UTC. Numbers (1–4) correspond to Leslie location in Fig. 2a.

$$RMSE = \sqrt{\frac{1}{N}\sum_{i=1}^{N}(Y_i - O_i)^2} \quad (1)$$

$$STDEV = \sqrt{\frac{1}{N}\sum_{i=1}^{N}(Y_i - \overline{Y})^2} \quad (2)$$

where $Y_i$ represents the model measurement, $O_i$ the observation, $\overline{Y}$ the measurement average and $N$ indicates every time step.

According to Evans and Guishard (2009), the concept of potential temperature in the dynamic tropopause ($\theta - PV$) is important to study the upper troposphere structure. The equivalence is observed using the vertical component of the Ertel, equation as follows:

$$PV = -\frac{g}{\rho}(\xi_\theta + f)\frac{\partial \theta}{\partial p} \quad (3)$$

where $\theta$, is the potential temperature, $\rho$, air density, $\xi_\theta$, relative vorticity $\xi_\theta = \partial_x v - \partial_y u$ in an isentropic surface, $f$ is the planetary vorticity and $\partial \theta / \partial p$ represents a stability parameter. Typically, PV is expressed in potential vorticity units (PVU), where $1\ PVU = 10^{-6} Km^2 kg^{-1} s^{-1}$.

The dynamic tropopause is defined as a region with a strong PV gradient that corresponds to the level of 2 PVU. This definition of tropopause allows us to attribute changes caused by atmospheric dynamics (Hoskins et al., 1985; Brennan et al., 2008; Bretherton, 1966) and, in the case of this study, to relate the $\theta - PV$ in 2 PVU to the PV associated with the synoptic systems. Positive (negative) $\theta - PV$ anomalies are associated with cyclonic (anticyclonic) relative vorticity and/or high (low) values of static stability (Hoskins et al., 1985).

Additionally, the geopotential height at 300 hPa ($Z$) is used in the current study as a tracer of the ridge-trough system; the 500–1000 hPa geopotential thickness ($\Delta Z$) is also considered to study the thermal configuration of the Leslie system. Moreover, absolute and relative differences in the fields of the ICs of the global models were also analyzed at 00:00 UTC on October 13.

### 3. Results and discussions

#### a. Trajectory, intensity, and synoptic environment

The trajectory and intensity behavior of tropical and extratropical systems are crucial parameters for decision-making purposes. In the case of the Leslie system, situated at a high latitude, its trajectory and intensity were influenced by the synoptic flow, particularly the interaction with an existent trough. Several atmospheric fields are employed in this study to identify differences in the system's initial position and the various configurations of synoptic flows surrounding the Leslie system.

On September 22, 2018, the precursors of Hurricane Leslie depict as an extratropical low located 1300 km western-southwestern of the Azores Islands. The extratropical low intensifies with hurricane-force winds and moves erratically. The system presents several transitions in its life cycle. This study is focused on the uncertainty associated with MPAS simulations in the ET presented in the final phase of Leslie's life, starting at 00:00 UTC on October 13 (black box in Fig. 1a) when the system reaches its peak intensity and begins its ET (Pasch and Roberts, 2019).

Both simulations show overestimation of Leslie's intensity; specifically, the MPAS-I intensifies the system to 164 km/h (Figs. 1d and 2a), corresponding to hurricane category 2 on the Saffir-Simpson scale (National Hurricane Center, 2012). Similarly, at 12:00 UTC on October 13, the MPAS-G intensifies the cyclone to 132 km/h (Figs. 1d and 2c), corresponding to a category 1 hurricane. The low ability of both simulations to reproduce the weakening of the system can be explained by the complex interaction of tropical systems moving to mid-latitudes and that interact with the jet stream (Chen et al., 2023; Anwender et al., 2008). In a TC, such a long phase of intensification is only possible if the ocean heat content is high and the vertical wind shear is low, as is typical of a barotropic atmosphere (López-Reyes and Peña, 2021). As of 15:00 UTC, MAPS-G shows a clear weakening of wind strength and a more zonal track to the east (Figs. 1b and 2c, d). On the other hand, the MPAS-I scenario keeps a constant wind speed of Leslie until 21:00 UTC (Fig. 1d). Thereafter, Leslie rapidly wakes following a northeastward trajectory (Figs. 1b and 2b).

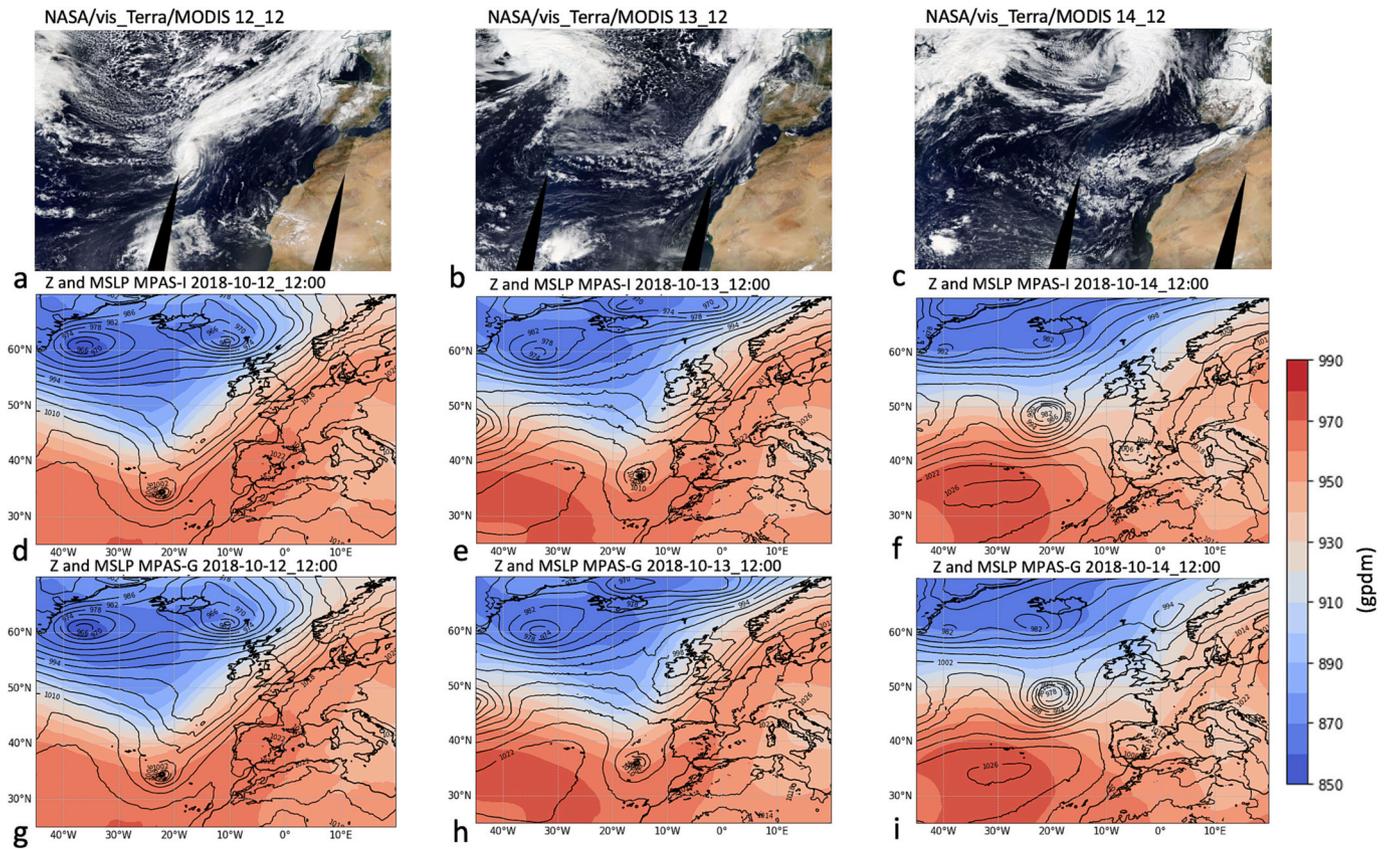

**Fig. 3.** Evolution of satellite imagery (True Colour: Red = Band I1, Green = Band M4, Blue = Band M3, from EOSDIS Worldview/MODIS) during the ET development on (a) 12 October, (b) 13 October, and (c) 14 October 2018, at 12:00 UTC. Source: EUMETSAT–Dundee Satellite Receiving Station (University of Dundee). Z at 300 hPa (shadow) and MSLP (contours) for: d-f) MPAS-I; g-i) MPAS-G simulations. (For interpretation of the references to colour in this figure legend, the reader is referred to the web version of this article.)

For the synoptic environment, both simulations show a wide trough embedding Leslie at 00:00 UTC on October 13. However, the MPAS-I (Figs. 3d-f) depicts a faster trough eastern displacement with Leslie, whereas Leslie remains at the edge of the trough axis and located southern in the case of the MPAS-G. Around 12:00 UTC on October 14, the MPAS-I displays a ridge at 300 hPa with a high pressure system at surface (outer isobar of 1022 hPa) very close to the Iberian Peninsula, forcing Leslie to move northwest (Fig. 3f). On the other hand, the extratropical system Michael, approximately located at 50°N, moves above the ridge on October 13 and 14. MPAS-G shows Michael's center further southern with a low-pressure center of 978 hPa (Figs. 3e, f), compared to 982 hPa of MPAS-I (Figs. 3h, i). This may have influenced the increase of MPAS-G trajectory errors because of the greater ridge amplitude where Michael accelerated the Leslie displacement as it interacted with the trough.

According to Keller et al. (2019) and Hoskins et al. (1985), potential temperature anomalies in the dynamic tropopause are associated with cyclonic PV anomalies, located on upper-level troughs. Although no anomalies are depicted in this study, the PV simulation differences (Figs. 4b-d) allow inferring some features of the trough that interacted with Leslie to be identified (see conceptual scheme in Fig. 4a). Figs. 4b-d show a band with negative θ − PV (positive) differences to the west (east) of the trough, which is consistent with the trough's slower displacement in the MPAS-I. In addition to being out of phase, the trough simulated by MPAS-I is less pronounced and displaced to the north (Fig. 4a). The above-mentioned suggests that the MAPS-G could be influenced from upstream flow (where Michael is), amplifying the ridge-trough (as in Figs. 4a) configuration and changing the behavior of the jet stream and therefore the trajectory of the Leslie system.

The evolution of the θ-PV and the ΔZ over the next few hours reveal a tendency to amplify (as a dipole) the differences between the model simulations (Figs. 5a, b). The positive ΔZ differences grow around Michael's position (on the ridge; western zone in Figs. 5a, b), that is, in MPAS-I, the synoptic flow around to Michael is shifted further to the north respect to MPAS-G. An obvious dipole with positive (negative) ΔZ differences of up to 85 gpm (70 gpm) north-northeast (west-southwest) of the trough axis at 24 h forecast (Figs. 5b) indicates the most pronounced configuration of the MPAS-I and the divergence in Leslie's trajectory in both simulations (eastern area in Fig. 5a, b). After 24 h, the difference in the Leslie center is 295 km. An error around 300 km is found (Fig. 1c) in the MPAS-G deviating from the recent trend of improvements in track forecasts reported by the NHC (Rappaport et al., 2009).

b. **Initial conditions analysis**

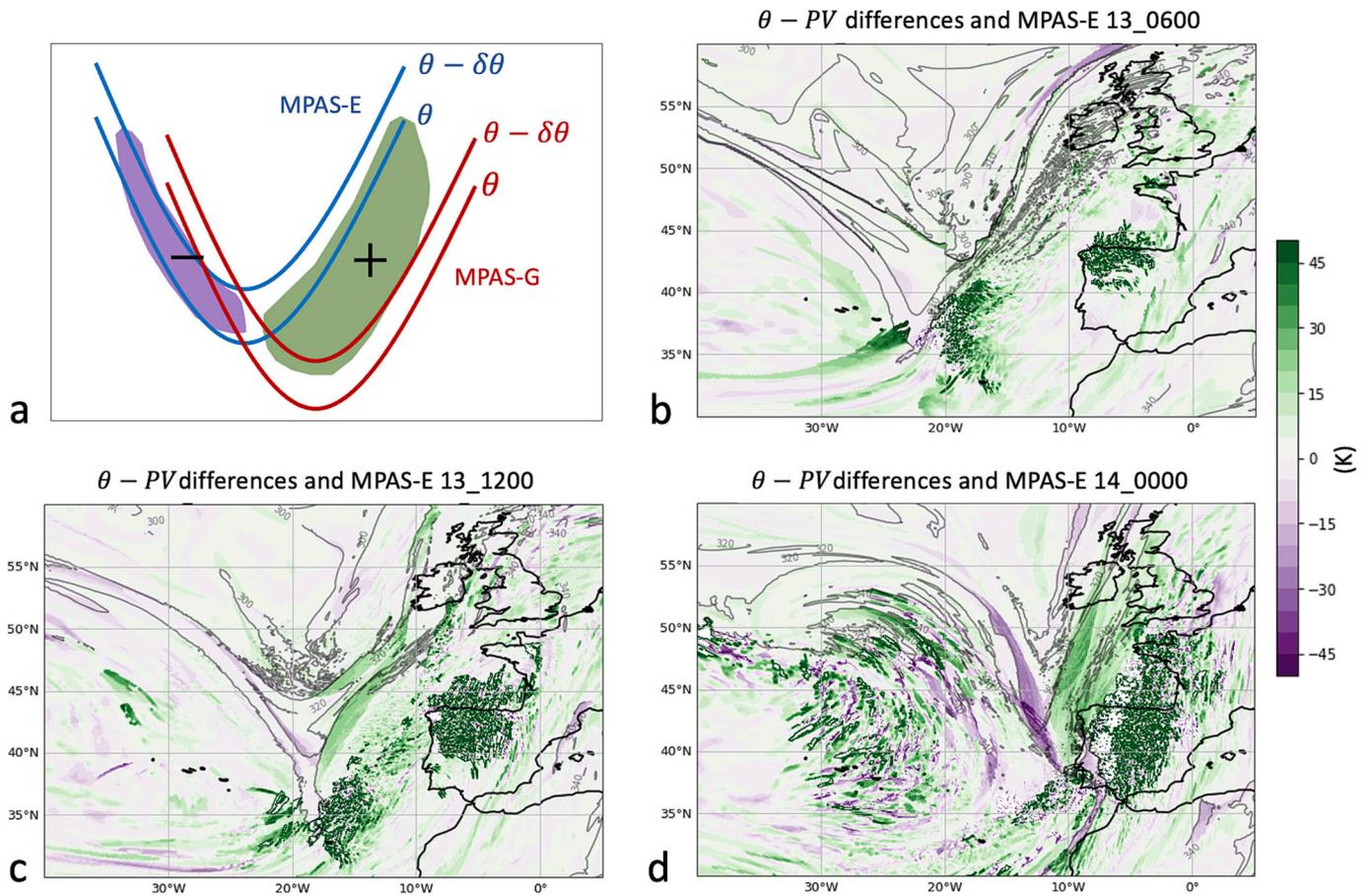

**Fig. 4.** (a) Conceptual scheme of the differences MPAS-I – MPAS-G (shadow) and isentropes (contours) for both simulations; $\theta - PV$ MPAS-I (contour) and differences of MPAS-I – MPAS-G (K; shadow) centered on Leslie system on: (b) 13 October at 06:00 UTC; (c) 13 October at 12:00 UTC and (d) 14 October at 00:00 UTC.

To identify the differences in the initial position of Leslie, as well as the differences in the flow and thermal characteristics of the environment in which the system is found, the IC fields of Z at 300 hPa, $\theta - PV$ and $\Delta Z$ are used. Considering a single MPAS configuration to simulate the Leslie, differences in trajectory should be only caused by the different global model analyses. The differences in the trajectory and intensity of the simulations are associated with the discrepancies in ICs and, possibly, amplified by the complex synoptic environment in which Leslie was evolving. On the one hand, the trajectory forecast for a system with ET depends mainly on the synoptic flow, while the intensity evolution is also related to the microphysical processes (Tao et al., 2011), as well as to the interaction of Leslie within the flow.

Over the mid-Atlantic ridge, a less intense dipole indicates smaller differences in the thermal configuration of the mid-troposphere between both model simulations (Figs. 5a, c). Michael, by then an extratropical cyclone, continues to ride on the ridge flow, slightly slower with IFS ICs, same for the case of Leslie. That is, the entire dorsal-trough synoptic system was delayed in MPAS-I compared to MPAS-G. Variations in the amplitude of the Rossby wave are also observed between the simulations (Figs. 5a, b). As mentioned in various studies (e.g., Komaromi and Doyle, 2018; Liu et al., 2012; Jones et al., 2003), the upstream PV adjustments caused by the diabatic processes in an ET (in this case with Michael) promote modifications on the structure of the ridge-trough configuration (both amplitude and intensity). Thus, the jet stream configuration is modified downstream, generating uncertainty in the interaction of Leslie with the trough and, therefore, in the trajectory of the system.

Fig. 6 shows the absolute differences (taking the IFS field as the reference) of the ICs between IFS and GFS for different fields. The dipole formed at 35°N, 20°W (Fig. 6 a) reveals as an indicator for the difference in the position of the Leslie center in the ICs. Positive (negative) 300 hPa Z values indicate the warm core of Leslie displaced to the north (south) in the IFS (GFS). Again, the greater $\Delta Z$ values located on the trough, above the mid-Atlantic ridge (Fig. 6b), indicate a higher intensity of the IFS values than the GFS ones, highlighting a weaker trough by the IFS and consequently a slightly lower mean temperature in middle-lower atmospheric levels on the Atlantic ridge.

Moreover, a weaker $\Delta Z$ dipole located at the Terranova coast (Figs. 6b) is associated with the post-ET of Hurricane Michael as an extratropical system. The dipole highlights a different cyclone position of Michael, promoting a different jet stream propagation as seen in Figs. 4b-d. According to Keller et al. (2019), Hulme and Martin (2009), and Liu et al. (2012), this initial flow modification triggers or amplifies a midlatitude Rossby wave packet, which disperses the impact of the ET into downstream regions, and it can contribute to increase the uncertainty in the behavior of the flow downstream, as it happened in the case of Leslie.

Concerning the differences in the $\theta - PV$ shown in Fig. 6c, the largest and most positive discrepancies in the ICs occur in the eastern area of the trough that interacted with Leslie. The differences between the models around Leslie are greater than +30 K (approximately 20%; Fig. S1a), suggesting that the MPAS-I keeps a higher absolute vorticity advection over the system and increased baroclinicity (Prezerakos et al., 1997). This results on a more intensity of the MPAS-I than the MPAS-G outputs. This is consistent with the MPAS outputs (Figs. 1d, 3a) for which the IFS overestimated the intensity of the cyclone throughout the selected period. An additional area with positive $\theta - PV$ differences around +20 K (approximately 15%; Fig. S1b) is located at Michael's upstream position (Fig. 6c). The MPAS-G dynamic tropopause, that serves as a flow motion guide, depicts differences related to such area about the position and motion of Michael and its dynamic interaction with the jet stream.

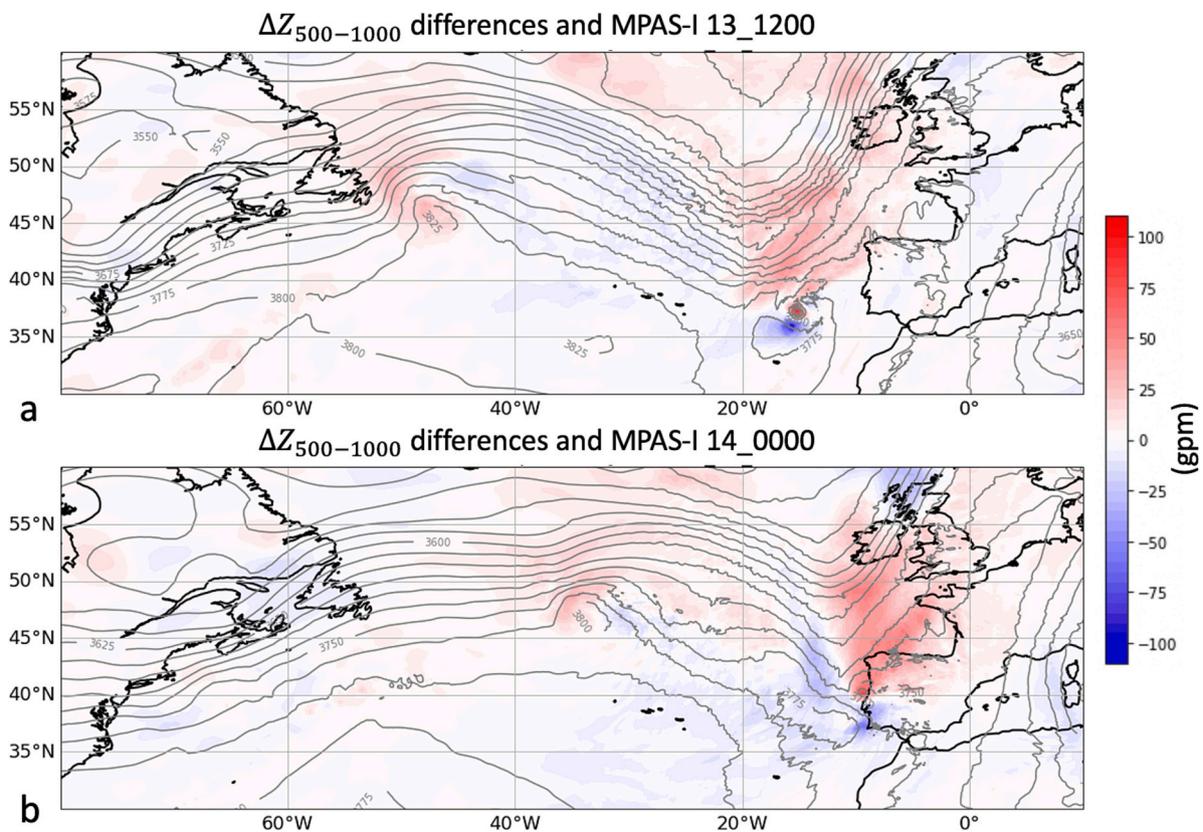

**Fig. 5.** $\Delta Z$ (gpm) of MPAS-I – MPAS-G (shadow) and MPAS-I (contour): (a) 13 October at 12:00 UTC and (b) 14 October at 00:00 UTC.

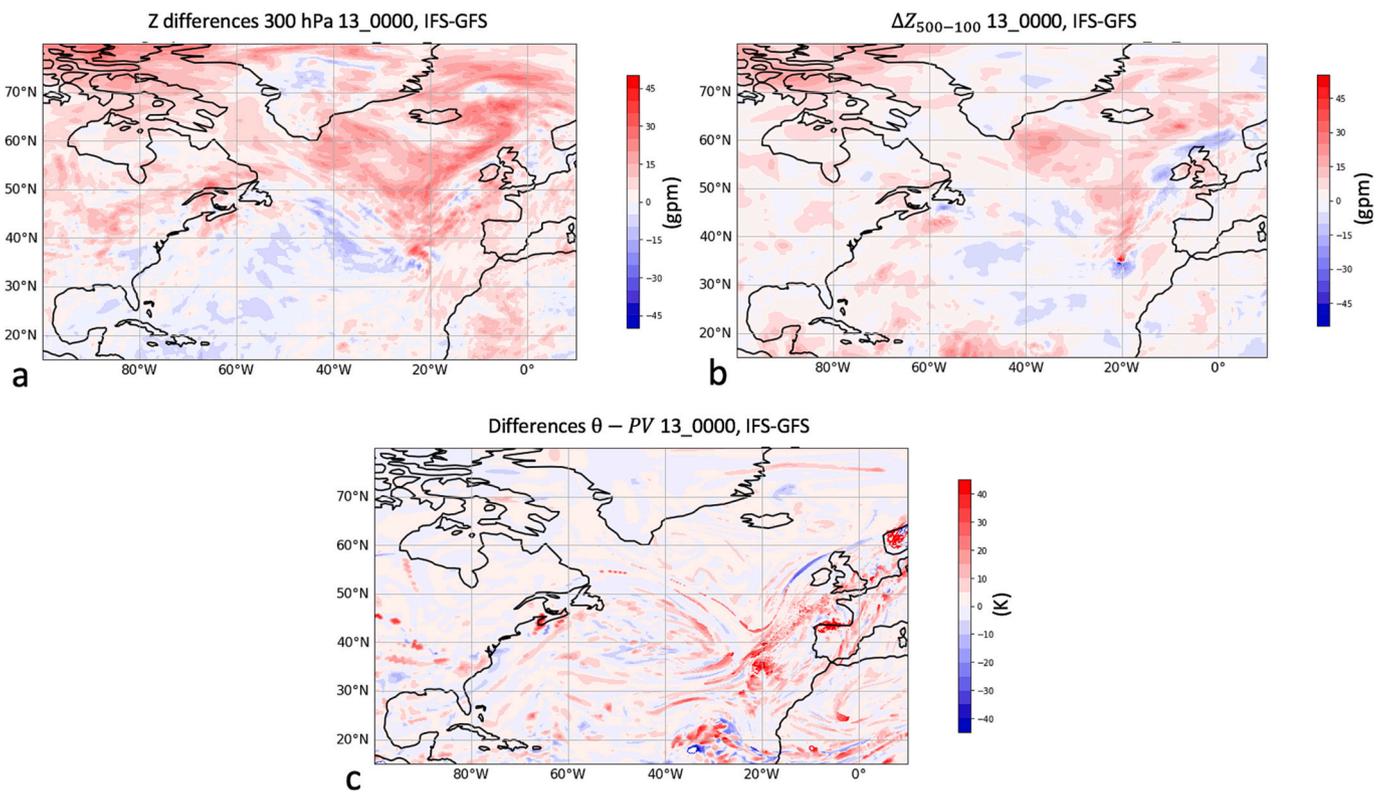

**Fig. 6.** (a) Absolute IC differences of $Z$ (gpm) at 300 hPa; (b) $\Delta Z$ (gpm) and (c) $\theta - PV$ (K).

The whole results show a better trajectory performance of the IFS ICs against the GFS ones. The generation of ICs is one of the fundamental parts to carry out a reliable simulation of the behavior of a meteorological system (Hamill et al., 2011; Zhang et al., 2019; Chen et al., 2022). In the current experiments, the analyzed fields are generated from the GFS and IFS operational data assimilation methods. For the GFS model, the data assimilation method is the Global Data Assimilation System (GDAS), while for the IFS the four-dimensional variational data assimilation formulation (4D-var) is used (details in Appendix 1). In both data assimilation methods, a field of analysis is generated by minimizing the cost function (Eq. A1 and A2), that is, the distance between the background field (also called first guess) and the observations (Kalnay, 2003). The initial conditions of 4D-var associated with IFS analysis can be better considered than GDAS, mainly due to the constant temporary corrections (see appendix A) made in the 4D-var with the observations to force the analyzes to more realistic scenarios. Additionally, a higher spatial resolution in the global domain and a largest quantity of observational data is considered in the 4D-var (European Center for Medium-Range Weather Forecasts, 2023a; National Center Environmental Prediction, 2021). On the other hand, the IFS has a greater capacity to represent convection, important in synoptic environments where convection is present as cumulus parameterizations can be a source of uncertainty both for the creation of the ICs, and for the forecast of trajectory and intensity of cyclones (Bassill, 2014; Bengtsson et al., 2019). Over the western and eastern ridge areas, where Michael and Leslie were located, respectively, the parameterization of clusters incorporated by GDAS may not accurately reality reproduce such configuration, affecting the $\theta - PV$ pattern and the subsequent downstream evolution. Finally, all data assimilation methods depend primarily on the quantity and quality of observations. Although in recent years the number of observations over the Atlantic Ocean has increased considerably, there are still regions with a lack of data at different vertical levels, which contributes to less precise interpolations by data assimilation methods which may affect the forecasts of trajectory and intensity of the models.

The importance of having good ICs for accurate modeling of a cyclone's trajectory is emphasized in this study, especially for high uncertainty systems as in the case of Hurricane Leslie. As found by Keller et al. (2019), upstream ET processes (as in the case of Hurricane Michael) could promote a source of anticyclonic PV, inducing downstream modifications in the Rossby waves amplitude, where Leslie is located. In the same way, the obtained wind intensity errors shown by both model outputs may be related to the complexity in simulating the interaction between Leslie and the upstream trough. Komaromi and Doyle, 2018 verified that the position of the cyclone with respect to the trough can result on a rapid and prolonged extratropical intensification (as happened in MPAS-I), particularly when the surface low pressure is located below a zone of divergence to the right of the trough axis.

## 4. Summary and conclusions

In this work, the impact of the ICs generated by the IFS and GFS global models to simulate the trajectory and intensity of the Leslie system with the MPAS model is analyzed. Two major findings emerge from the simulations: 1) the ability of each simulation to accurately replicate the trajectory and intensity of the Leslie system during its ET, and 2) the evaluation the impact of the generation of precise ICs for correctly modeling situations of high uncertainty, such as in the case study presented.

To sum up, the key points of this study are as follows:

1. The MPAS-I simulation outperforms MPAS-G in reproducing the system trajectory, with an error of <60 km, 24 h after the start of the simulation. Conversely, MPAS-G exhibits increasing errors when Leslie's center makes landfall (295 km at 03:00 UTC on 14 October). From a dynamic point of view, the better performance of MPAS-I is partly due to the better simulation of the upper flow where Leslie was embedded (on the trough axis). Also, this simulation may have better resolved Michael's upstream influence on the ridge. This should be verified in further studies by running simulations with different configurations for Michael, checking the relevance of the PV generation in the uncertainty of the evolution of the downstream flow.

2. A more intense ridge in the MPAS-I simulation is quite revealed, especially in the $\theta - PV$ field, that is, the jet stream configuration with which the Leslie system is interacting, is modified by Michael's upstream interaction. This distorts the trough structure and facilitates Leslie's channeling and subsequent movement within the jet flow further northern, just as really happened.

3. Both IC configurations are not able to satisfactorily model the intensity of the system, especially MPAS-I simulation. Throughout the MPAS-I simulation, the intensity is overestimated, even showing Leslie as a hurricane with winds corresponding to hurricane category 2, when the system was in an ET and constant weakening. We believe that, as demonstrated by Komaromi and Doyle (2018), the complex interaction between a cyclone and a trough can intensify the cyclone due to upper-level divergence wind shear values are not as high in the cyclone's vicinity.

4. In both simulations, particularly in MPAS-I, the Leslie location is placed in a region of upper-level divergence, favoring upward vertical movements and intensification of the system. This is one of the reasons why using accurate ICs is important to generating good cyclone forecasts, particularly during ET.

5. Finally, based on the results of this study, it is assumed that Leslie's ET is conditioned by the uncertainty of the ICs coming from the upstream flow where Michael is also an ET. The results of the experiments show the great importance of generating accurate ICs even after a few hours of simulation. The clear advantage of the IFS 4D-var as a data assimilation method resides in the greater number of used observations, the greater spatial resolution, and the constant correction of the analysis fields from the assimilated observations, improving the precision of the ICs. This allows simulations and even operational forecasts to be closer to reality under complex synoptic situations, as in the case of this study.

The number of tropical systems that have reached the western European zone is increasing in recent years. Therefore, the study of these events, particularly those that experience several transitions during their life cycle, reveals as necessary the use of accurate simulations with numerical weather prediction models to give a helpful tool to the forecasters when a more certain prediction of such events is needed.

**CRediT authorship contribution statement**

**M. López-Reyes:** Conceptualization, Methodology, Software, Formal analysis, Writing – original draft. **J.J. Gónzalez-Alemán:** Conceptualization, Methodology, Formal analysis. **M. Sastre:** Methodology, Supervision, Writing – review & editing. **D. Insua-Costa:** Conceptualization, Methodology, Formal analysis, Software. **P. Bolgiani:** Methodology, Writing – review & editing, Supervision. **M.L. Martín:** Conceptualization, Methodology, Writing – review & editing, Supervision, Funding acquisition.

**Declaration of Competing Interest**

The authors declare that they have no known competing financial interests or personal relationships that could have appeared to influence the work reported in this paper.

**Data availability**

Data will be made available on request.

**Acknowledgments**

This work was partially supported by the research project PID2019-105306RB-I00/AEI/10.13039/501100011033, and the two ECMWF Special Projects (SPESMART and SPESVALE). We would like to thank Instituto Frontera A.C. for financial support to carry out the work. MPAS simulations were run by DIC at CESGA (Centro de Supercomputación de Galicia), Santiago de Compostela, Galicia, Spain.

**Appendix A. Supplementary data**

Supplementary data to this article can be found online at https://doi.org/10.1016/j.atmosres.2023.107020.

**References**

Anwender, D., Harr, P.A., y Jones, S.C., 2008. Previsibilidad asociada con los impactos aguas abajo de la transición extratropical de los ciclones tropicales: estudios de casos. Mon. Weather Rev. 136 (9), 3226–3247. https://doi.org/10.1175/2008MWR2249.1.

Bassill, N.P., 2014. Accuracy of early GFS and ECMWF Sandy (2012) track forecasts: evidence for a dependence on cumulus parameterization. Geophys. Res. Lett. 41 (9), 3274–3281. https://doi.org/10.1002/2014GL059839.

Bengtsson, L., Dias, J., Gehne, M., Bechtold, P., Whitaker, J., Bao, J., Magnusson, L., Michelson, S., Pegion, P., Tulich, S., Kiladis, G.N., 2019. Convectively coupled Equatorial Wave Simulations using the ECMWF IFS and the NOAA GFS Cumulus Convection Schemes in the NOAA GFS Model. Mon. Weather Rev. 147 (11), 4005–4025. https://doi.org/10.1175/MWR-D-19-0195.1.

Beven II, J.L., Berg, R., Hagen, A., 2019. Hurricane Michael (AL142018). National Hurricane Center. Tropical Cyclone Report.

Bieli, M., Camargo, S.J., Sobel, A.H., Evans, J.L., Hall, T., 2019. A Global Climatology of Extratropical transition. Part I: Characteristics across Basins. J. Clim. 32 (12), 3557–3582. https://doi.org/10.1175/JCLI-D-17-0518.1.

Brennan, M.J., Lackmann, G.M., Mahoney, K.M., 2008. Potential Vorticity (PV) Thinking in Operations: the Utility of Nonconservation. Weather Forecast. 23 (1), 168–182. https://doi.org/10.1175/2007WAF2006044.1.

Bretherton, F.P., 1966. Critical layer instability in baroclinic flows. Q. J. R. Meteorol. Soc. 92 (393), 325–334.

Chen, S., Zhang, Y., Xu, J., Shen, W., Ye, G., Lu, Z., 2022. Data assimilation of adaptive observation and application for typhoon forecasts over the Western North Pacific. Atmos. Res. 276, 106274. https://doi.org/10.1016/j.atmosres.2022.106274.

Chen, G., Ling, J., Lin, Z., Xiao, Z., Li, C., 2023. Role of tropical-extratropical interactions in the unprecedented 2022 extreme rainfall in Pakistan: a historical perspective. Atmos. Res. 291, 106817. https://doi.org/10.1016/j.atmosres.2023.106817.

Duda, M., Laura, Fowler, Skamarock, B., Roesch, C., Jacobsen, D., Ringler, T., 2019. MPAS-Atmosphere Model User's Guide (version 7.0). https://www2.mmm.ucar.edu/projects/mpas/mpas_atmosphere_users_guide_7.0.pdf.

European Center for Medium-Range Weather Forecasts, 2023a. Forecast User Portal. https://confluence.ecmwf.int/display/FCST/Forecast+User+Portal.

European Center for Medium-Range Weather Forecasts, 2023b. Implementation of IFS Cycle 45r1. Changes to the Forecasting System. https://confluence.ecmwf.int/display/FCST/Implementation+of+IFS+cycle+45r1.

Evans, J.L., Guishard, M.P., 2009. Atlantic subtropical storms. Part I: diagnostic criteria and composite analysis. Mon. Weather Rev. 137 (7), 2065–2080. https://doi.org/10.1175/2009MWR2468.1.

Evans, C., Wood, K.M., Aberson, S.D., Archambault, H.M., Milrad, S.M., Bosart, L.F., Corbosiero, K.L., Davis, C.A., Dias Pinto, J.R., Doyle, J., Fogarty, C., Galarneau Jr., T.J., Grams, C.M., Griffin, K.S., Gyakum, J., Hart, R.E., Kitabatake, N., Lentink, H.S., McTaggart-Cowan, R., Perrie, W., Quinting, J.F.D., Reynolds, C.A., Riemer, M., Ritchie, E.A., Sun, Y., Zhang, F., 2017. The extratropical transition of tropical cyclones. Part I: cyclone evolution and direct impacts. Mon. Weather Rev. 145 (11), 4317–4344. https://doi.org/10.1175/MWR-D-17-0027.1.

Hamill, T.M., Whitaker, J.S., Kleist, D.T., Fiorino, M., Benjamin, S.G., 2011. Predictions of 2010's tropical cyclones using the GFS and ensemble-based data assimilation methods. Mon. Weather Rev. 139 (10), 3243–3247.

Hoskins, B.J., McIntyre, M.E., Robertson, A.W., 1985. On the use and significance of isentropic potential vorticity maps. Q.J.R. Meteorol. Soc. 111, 877–946. https://doi.org/10.1002/qj.49711147002.

Hulme, A.L., Martin, J.E., 2009. Synoptic- and Frontal-Scale Influences on Tropical transition events in the Atlantic Basin. Part I: A Six-Case Survey. Mon. Weather Rev. 137 (11), 3605–3625. https://doi.org/10.1175/2009MWR2802.1.

Jones, S.C., Harr, P.A., Abraham, J., Bosart, L.F., Bowyer, P.J., Evans, J.L., Hanley, D.E., Hanstrum, B.N., Hart, R.E., Lalaurette, F., Sinclair, M.R., Smith, R.K., Thorncroft, C., 2003. The Extratropical Transition of Tropical Cyclones: Forecast Challenges, Current Understanding, and Future Directions, 18 (6), 1052–1092. https://doi.org/10.1175/1520-0434(2003)018<1052:TETOTC>2.0.CO;2.

Kalnay, E., 2003. Atmospheric Modeling, Data Assimilation and Predictability. Cambridge university Press.

Keller, J.H., Grams, C.M., Riemer, M., Archambault, H.M., Bosart, L., Doyle, J.D., Evans, J.L., Galarneau Jr., T.J., Griffin, K., Harr, P.A., Kitabatake, N., McTaggart-Cowan, R., Pantillon, F., Quinting, J.F., Reynolds, C.A., Ritchie, E.A., Torn, R.D., Zhang, F., 2019. The Extratropical transition of Tropical Cyclones. Part II: Interaction with the Midlatitude Flow, Downstream Impacts, and Implications for Predictability. Mon. Weather Rev. 147 (4), 1077–1106. https://doi.org/10.1175/MWR-D-17-0329.1.

Komaromi, W.A., Doyle, J.D., 2018. Sobre la dinámica de las interacciones entre ciclones tropicales y canales. J. Atmos. Sci. 75 (8), 2687–2709. https://doi.org/10.1175/JAS-D-17-0272.1.

Liu, L., Davidson, N.E., Zhu, H., Lok, C.C., 2012. Downstream development during the extra- tropical transition of tropical cyclones: Observational evidence and influence on storm structure. Trop. Cyclone Res. Rev. 1 (4), 430–447. https://doi.org/10.6057/2012TCRR04.02.

López-Reyes, M., Peña, Á.R.M., 2021. Comparación de las variables físicas que influyen en la rÁpida intensificación de los ciclones tropicales del Océano Pacífico nororiental durante el periodo 1970–2018. Cuadernos GeogrÁficos 60 (2), 105–125. https://doi.org/10.30827/cuadgeo.v60i2.15474.

Maddison, J.W., Gray, S.L., Martínez-Alvarado, O., Williams, K.D., 2020. Impact of model upgrades on diabatic processes in extratropical cyclones and downstream forecast evolution. Q. J. R. Meteorol. Soc. 146 (728), 1322–1350. https://doi.org/10.1002/qj.3739.

Mandement, M., Caumont, O., 2021. A numerical study to investigate the roles of former Hurricane Leslie, orography and evaporative cooling in the 2018 Aude heavy-precipitation event. Weather and Climate Dynam. 2 (3), 795–818.

McTaggart-Cowan, R., Gyakum, J.R., Yau, M.K., 2003. The influence of the downstream state on extratropical transition: Hurricane Earl (1998) case study. Mon. Weather Rev. 131, 1910–1929.

McTaggart-Cowan, R., Gyakum, J.R., Yau, M.K., 2004. The Impact of Tropical Remnants on Extratropical Cyclogenesis: Case Study of Hurricanes Danielle and Earl (1998). Mon. Weather Rev. 132 (8), 1933–1951 doi:10.1175/1520-0493(2004)132<1933:TIOTRO>2.0.CO;2.

Munsell, E.B., Zhang, F., 2014. Prediction and uncertainty of Hurricane Sandy (2012) explored through a real-time cloud-permitting ensemble analysis and forecast system assimilating airborne Doppler radar observations. J. Adv. Model. Earth Syst. 6 (1), 38–58. https://doi.org/10.1002/2013MS000297.

Munsell, E.B., Sippel, J.A., Braun, S.A., Weng, Y., Zhang, F., 2015. Dynamics and predictability of Hurricane Nadine (2012) evaluated through convection-permitting ensemble analysis and forecasts. Mon. Weather Rev. 143 (11), 4514–4532. https://doi.org/10.1175/MWR-D-14-00358.1.

National Center Environment Information, 2023. The Global Forecast System. https://www.emc.ncep.noaa.gov/emc/pages/numerical_forecast_systems/gfs.php.

National Center Environmental Prediction, 2021. List of GFS Implementations. https://www.emc.ncep.noaa.gov/emc/pages/numerical_forecast_systems/gfs/implementations.php.

National Hurricane Center, 2012. Saffir-Simpson Hurricane Wind Scale. https://www.nhc.noaa.gov/aboutsshws.php.

Pantillon, F.P., Chaboureau, J., Mascart, P.J., Lac, C., 2013. Predictability of a Mediterranean Tropical-like storm Downstream of the Extratropical transition of Hurricane Helene (2006). Mon. Weather Rev. 141 (6), 1943–1962. https://doi.org/10.1175/MWR-D-12-00164.1ç.

Pasch, R.J., Roberts, D.P., 2019. Hurricane Leslie (AL132018). National Hurricane Center. Tropical Cyclone Report.

Pérez-Alarcón, A., Fernández-Alvarez, J.C., Sorí, R., Liberato, M.L., Trigo, R.M., Nieto, R., Gimeno, L., 2023. How much of precipitation over the Euroregion Galicia–Northern Portugal is due to tropical-origin cyclones?: a Lagrangian approach. Atmos. Res. 106640 https://doi.org/10.1016/j.atmosres.2023.106640.

Prezerakos, N.G., Flocas, H.A., Michaelides, S.C., 1997. Absolute vorticity advection and potential vorticity of the free troposphere as synthetic tools for the diagnosis and forecasting of cyclogenesis. Atmosphere-ocean 35 (1), 65–91. https://doi.org/10.1080/07055900.1997.9649585.

Quitián-Hernández, L., Martín, M.L., González-Alemán, J.J., Santos-Muñoz, D., Valero, F., 2016. Identification of a subtropical cyclone in the proximity of the Canary Islands and its analysis by numerical modeling. Atmos. Res. 178, 125–137. https://doi.org/10.1016/j.atmosres.2016.03.008.

Rappaport, E.N., Franklin, J.L., Avila, L.A., Baig, S.R., Beven II, J.L., Blake, E.S., Burr, C.A., Jiing, J., Juckins, C.A., Knabb, R.D., Landsea, C.W., Mainelli, M., May-Field, M., McAdie, C.J., Pasch, R.J., Sisko, C., Stewart, S.R., Tribble, A.N., 2009. Advances and challenges at the national hurricane center. Weather Forecast. 24 (2), 395–419. Retrieved Jun 6, 2022, from. https://journals.ametsoc.org/view/journals/wefo/24/2/2008waf222212.

Rauscher, S.A., Ringler, T.D., 2014. Impact of Variable-Resolution Meshes on Midlatitude Baroclinic Eddies using CAM-MPAS-A. Mon. Weather Rev. 142 (11), 4256–4268. Retrieved Nov 15, 2021, from. https://journals.ametsoc.org/view/journals/mwr/142/11/mwr-d-13-00366.1.Xml. https://doi.org/10.1175/MWR-D-13-00366.1.

Sadler, J.C., 1976. A Role of the Tropical Upper Tropospheric Trough in early season Typhoon Development. Mon. Weather Rev. 104 (10), 1266–1278. https://doi.org/10.1175/1520-0493(1976)104<1266:AROTTU>2.0.CO;2.

Scheck, L., Jones, S.C., Juckes, M., 2011. The Resonant Interaction of a Tropical Cyclone and a Tropopause Front in a Barotropic Model. Part I: Zonally Oriented Front. J. Atmos. Sci. 68 (3), 405–419. https://doi.org/10.1175/2010JAS3482.1.

Skamarock, W.C., Klemp, J.B., Duda, M.G., Fowler, L.D., Park, S.H., Ringler, T.D., 2012. A multiscale nonhydrostatic atmospheric model using centroidal Voronoi tesselations and C-grid staggering. Mon. Weather Rev. 140 (9), 3090–3105.

Skamarock, W.C., Klemp, J.B., Dudhia, J., Gill, D.O., Barker, D., Duda, M.G., Powers, J. G., 2008. A Description of the Advanced Research WRF Version 3 (no. NCAR/TN-475+STR). University Corporation for Atmospheric Research. https://doi.org/10.5065/D68S4MVH.

Stewart, S., 2018. BULLETIN Post-Tropical Cyclone Leslie Advisory Number, 70. Available at: https://www.nhc.noaa.gov/archive/2018/al13/al132018.public.070.shtml?.

Tao, W.K., Shi, J.J., Chen, S.S., Lang, S., Lin, P.L., Hong, S.Y., Hou, A., 2011. The impact of microphysical schemes on hurricane intensity and track. Asia-Pacif. J. Atmos Sci. 47, 1–16. https://doi.org/10.1007/s13143-011-1001-z.

Torn, R.D., Whitaker, J.S., Pegion, P., Hamill, T.M., Hakim, G.J., 2015. Diagnosis of the source of GFS Medium-Range Track Errors in Hurricane Sandy (2012). Mon. Weather Rev. 143 (1), 132–152. https://doi.org/10.1175/MWR-D-14-00086.1.

Veren, D., Evans, J.L., Jones, S., Chiaromonte, F., 2009. Novel metrics for evaluation of ensemble forecasts of tropical cyclone structure. Mon. Weather Rev. 137 (9), 2830–2850.

Zhang, L., Liu, Y., Liu, Y., Gong, J., Lu, H., Jin, Z., Zhao, B., 2019. The operational global four-dimensional variational data assimilation system at the China Meteorological Administration. Q. J. R. Meteorol. Soc. 145 (722), 1882–1896. https://doi.org/10.1002/qj.3533.